%% file: paper.tex
\documentclass[10pt,conference]{IEEEtran}
\IEEEoverridecommandlockouts
\usepackage{cite}
\usepackage{amsmath,amssymb,amsfonts}
\usepackage{algorithmic}
\usepackage{graphicx}
\usepackage{textcomp}
\usepackage{xcolor}
\usepackage{balance}
\usepackage[hidelinks]{hyperref}
\usepackage[capitalise]{cleveref}
\usepackage[all]{nowidow}
\usepackage[caption=false]{subfig}
\usepackage{booktabs}
\usepackage{tabularx}
\usepackage{dblfloatfix}

\begin{document}

\title{Streaming vs. Functions:\\ A Cost Perspective on Cloud Event Processing}

\author{\IEEEauthorblockN{Tobias Pfandzelter\IEEEauthorrefmark{2}\IEEEauthorrefmark{1}, S\"oren Henning\IEEEauthorrefmark{3}\IEEEauthorrefmark{1}, Trever Schirmer\IEEEauthorrefmark{2}, Wilhelm Hasselbring\IEEEauthorrefmark{3}, David Bermbach\IEEEauthorrefmark{2}}\thanks{\IEEEauthorrefmark{1}Co-first authors.}
    \IEEEauthorblockA{\IEEEauthorrefmark{2}\textit{TU Berlin \& ECDF, Mobile Cloud Computing Research Group}\\
        \{tp,ts,db\}@mcc.tu-berlin.de}
    \IEEEauthorblockA{\IEEEauthorrefmark{3}\textit{Kiel University, Software Engineering Group} \\
        \{soeren.henning,hasselbring\}@email.uni-kiel.de}

}

\maketitle

\begin{abstract}
    In cloud event processing, data generated at the edge is processed in real-time by cloud resources.
    Both distributed stream processing (DSP) and Function-as-a-Service (FaaS) have been proposed to implement such event processing applications.
    FaaS emphasizes fast development and easy operation, while DSP emphasizes efficient handling of large data volumes.
    Despite their architectural differences, both can be used to model and implement loosely-coupled job graphs.

    In this paper, we consider the selection of FaaS and DSP from a cost perspective.
    We implement stateless and stateful workflows from the Theodolite benchmarking suite using cloud FaaS and DSP.
    In an extensive evaluation, we show how application type, cloud service provider, and runtime environment can influence the cost of application deployments and derive decision guidelines for cloud engineers.
\end{abstract}


\input{sections/1_introduction.tex}

\input{sections/2_background.tex}

\input{sections/3_benchmark.tex}
\input{sections/4_experiments.tex}

\input{sections/5_guidelines.tex}

\input{sections/6_discussion.tex}

\input{sections/7_relwork.tex}

\input{sections/8_conclusion.tex}

\section*{Acknowledgements}

Partially funded by the Deutsche Forschungsgemeinschaft (DFG, German Research Foundation) -- 415899119.
This material is based upon works supported by the Google Cloud Research Credits program with the awards GCP209186206 and GCP203304083.

\balance

\bibliographystyle{IEEEtran}
\bibliography{bibliography}

\end{document}

%% file: sections/1_introduction.tex
\section{Introduction}
\label{sec:introduction}

The increasing degree of data generation at the edge, e.g., by web clients or IoT devices, has led to a growing demand for live data and event processing in the cloud~\cite{Thamsen2022,paper_pfandzelter_zero2fog,paper_bermbach_cloud_engineering}.
Today, the most popular paradigms for this are \emph{distributed stream processing} (DSP) and \emph{Function-as-a-Service} (FaaS)~\cite{BDR2021,paper_pfandzelter_functions_streams}.
In both paradigms, data processing applications are modeled as loosely-coupled graphs of data operations.%

In DSP, this graph is a network of \emph{operators}, deployed on a stream processing engine running on a distributed cluster of compute nodes.
The stream processing engine partitions incoming data across nodes for horizontal scalability, hence, parallelizing the data processing workflow for the developer~\cite{Thamsen2022,Margara2022}.
Typical examples of stream processing engines include Apache Flink\footnote{\url{https://flink.apache.org/}} and Google Cloud Dataflow\footnote{\url{https://cloud.google.com/dataflow/}}~\cite{Fragkoulis2020,Akidau2021}.
FaaS platforms, e.g., AWS Lambda\footnote{\url{https://aws.amazon.com/lambda/}} and Google Cloud Functions\footnote{\url{https://cloud.google.com/functions/}}, allow developers to deploy small, stateless functions on managed infrastructure that are billed per invocation and run duration.
These functions can also be chained to build larger applications, e.g., through synchronous invocations or asynchronously by sharing state in a database~\cite{Mahgoub2021-dn,Copik2022-sv}.
The managed approach promises high elasticity and scalability for developers and allows cloud service providers to allocate their infrastructure more efficiently~\cite{Eismann2021,paper_hendrickson_openlambda}.

Despite their architectural differences, both DSP and FaaS can be used to model the loosely-coupled job graphs that underlie cloud data processing~\cite{Castro2019,paper_pfandzelter_functions_streams}.
Beyond some qualitative concerns, different billing models introduce a cost dimension that should be taken into account when designing data processing applications and choosing between paradigms~\cite{Van_Eyk2018-vc,Eismann2021}.
In this paper, we quantify this cost dimension through cost benchmarking~\cite{book_cloud_service_benchmarking} to let application developers and cloud engineers make more informed decisions when designing event processing applications.
We make the following contributions:

\begin{itemize}
    \item We present an application-centric benchmark with both stateful and stateless applications for cost-benchmarking of DSP and FaaS environments (\cref{sec:benchmark}).
    \item In experiments, we analyze the impact of processing paradigm, type of application, execution environment, and choice of cloud provider on the cost of an event processing application deployment (\cref{sec:experiments}).
    \item We provide decision guidelines for application developers based on our quantitative data (\cref{sec:guidelines}).
    \item We discuss the limitations of our work and derive avenues for future work (\cref{sec:discussion}).
\end{itemize}

We make our implementation available as open-source\footnote{\url{https://github.com/pfandzelter/cloud-event-processing-costs}} to enable other researchers and practitioners to conduct their own experiments.

%% file: sections/2_background.tex
\section{Background}
\label{sec:background}

While the concept of cloud computing is well-established in both research and industry, paradigms for cloud applications are constantly evolving.
In this section, we give an overview of distributed stream processing and Functions-as-a-Service, two of today's most common cloud data processing paradigms~\cite{Thamsen2022,paper_pfandzelter_functions_streams}, and introduce the related terminology.

\subsection{Distributed Stream Processing}

Most distributed stream processing engines extend the well-known MapReduce pattern~\cite{DeanGhemawat2010} with support for processing continuous data streams.
In modern DSP engines, developers define dataflow graphs (called \textit{pipelines} or \textit{jobs}) of \textit{operators} using a declarative programming model~\cite{Fragkoulis2020, Margara2022}.
Prominent examples for DSP engines are the open source projects Apache Flink~\cite{Carbone2015}, Apache Samza~\cite{Noghabi2017}, and Apache Kafka Streams~\cite{Wang2021}, or cloud services such as Google Cloud Dataflow~\cite{Akidau2015}. Apache Beam\footnote{\url{https://beam.apache.org/}} is a framework providing a unified programming model~\cite{Akidau2015} to define dataflow graphs, which can be executed by many stream processing engines.

DSP engines are deployed as clusters of multiple instances (e.g., on different computing nodes).
To enable horizontal scalability, data streams between operators are partitioned and operators are scheduled on multiple instances, where each operator instance processes only a portion of the data.
The key idea is that state should only be maintained locally in an operator instance.
Additionally, DSP engines often use periodic checkpointing and require durable, replayable data sources to ensure fault tolerance.

While stream processing engines have traditionally been operated as long-running clusters on virtual machines, they are now often deployed in standalone, cloud-native applications. In particular, containerization techniques and Kubernetes, the de-facto standard for container orchestration~\cite{CNCF2022}, are used to reduce the operational complexity when running DSP jobs at large scale; managed Kubernetes services are provided on all major cloud platforms.
In addition to a (fixed) cluster management fee, users of such services are billed variable cost for the allocated VMs or, more recently, for the actual resource usage of containers.

\subsection{Function-as-a-Service}

In the FaaS programming model, developers deploy applications in the form of individual functions to a FaaS platform that handles event-driven code invocation and horizontal scaling.
Function infrastructure is completely handled by the cloud service provider, i.e., ``serverless'', and consumers pay per request based on the resources consumed by a function~\cite{paper_hendrickson_openlambda,Scheuner2020-wr}.
Functions can be implemented in a number of programming languages and can be invoked by web requests, IoT sensor readings, database updates, and even other functions, so-called function chaining~\cite{Jia2021-nn,paper_grambow_befaas}.

A key element to horizontal scalability is that function instances logically exist only for the duration of a single invocation and do not support any state beyond that execution~\cite{Copik2022-sv}.
To support stateful applications, functions usually leverage serverless, pay-per-request cloud datastores such as Google Cloud Firestore\footnote{\url{https://cloud.google.com/firestore/}} or AWS DynamoDB\footnote{\url{https://aws.amazon.com/dynamodb/}}~\cite{Akhter2019,Sreekanti2020}.

In combination with lightweight virtualization techniques, such as containers or microVMs~\cite{paper_pfandzelter_tinyfaas,Agache2020-ug}, FaaS platforms can quickly spin up new and destroy old function instances, enabling rapid elasticity.
The low management burden for the consumer and the wide range of possible applications are clear advantages for developers.
For cloud service providers, the fine-grained execution of functions enables a more efficient allocation of their infrastructure~\cite{paper_hendrickson_openlambda}.

%% file: sections/3_benchmark.tex
\section{Cost Benchmark}
\label{sec:benchmark}

Both FaaS and DSP can be used to build cloud event processing applications.
To quantify the cost dimension of the decision between the two paradigms when building such an application, we introduce a new application-centric cost benchmark that can be applied to any cloud processing paradigm.
The proposed benchmark comprises an application implementing two example use-cases, which could easily be extended, the load generator which creates requests for the application, and its configuration.
The system under test (SUT) in each benchmark is either a FaaS or a DSP platform.
In this section, we present our proposed benchmark and the methodology for executing it.

\subsection{Cloud Event Processing Use-Case}

Our example application is derived from the Theodolite suite of stream processing scalability benchmarks~\cite{BDR2021}.
Both use-cases are designed for Industrial Internet of Things (IIoT) event processing in the context of a smart factory, where sensors at the edge produce large amounts of data that require real-time event processing in the cloud~\cite{DIMA2021}.
We chose both a stateless and a stateful use-case in order to quantify the impact of state management on application cost.

\subsubsection*{\textbf{Stateless Storage}}
\begin{figure}
    \centering
    \includegraphics[height=1cm]{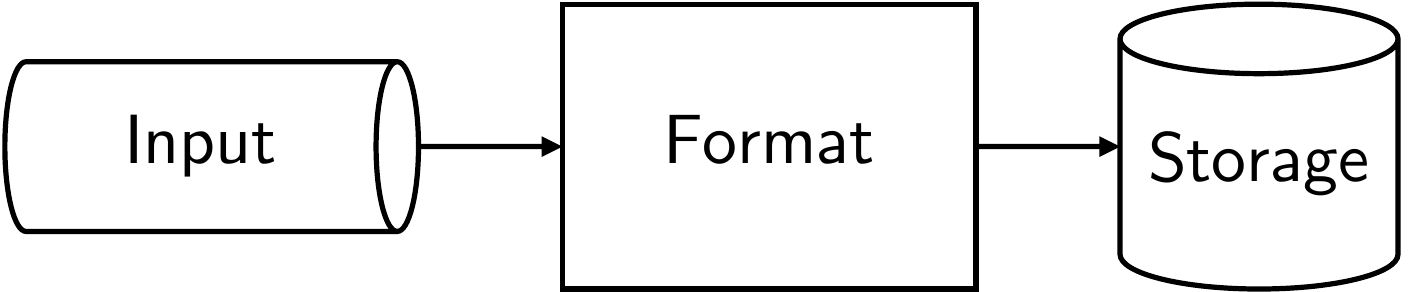}
    \caption{In the stateless storage use-case (\emph{UC1}), input data is transformed and then persisted in a storage backend.}
    \label{fig:uc1}
\end{figure}
The first use-case (\emph{UC1}) persists incoming data in a serverless key-value database.
Such an operation is often required for archiving events and making them accessible to other applications.
As shown in \cref{fig:uc1}, incoming events are first transformed to match the data format, required by the database API, and then written to that database system.
Since each data item is treated individually, no state is maintained within the application.

\subsubsection*{\textbf{Stateful Sliding Window Aggregation}}
\begin{figure}
    \centering
    \includegraphics[height=1cm]{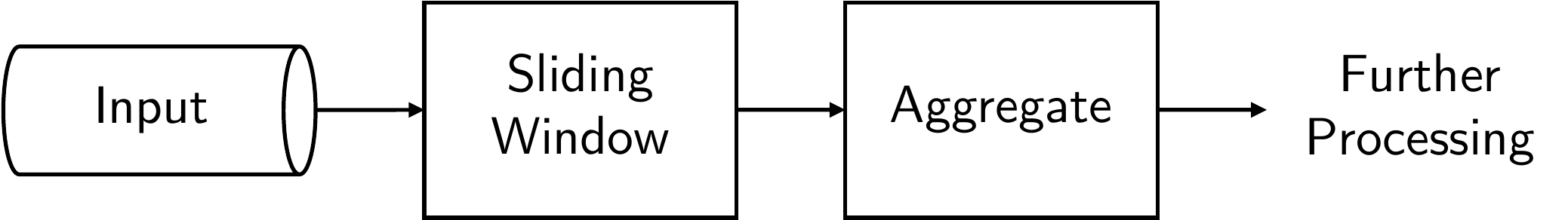}
    \caption{In the stateful sliding window aggregation use-case (\emph{UC2}), incoming data items are grouped in fixed-size time windows based on their key. Within a window, data is aggregated and then forwarded for further processing, e.g., to persist it.}
    \label{fig:uc3}
\end{figure}

\begin{table*}
    \centering
    \caption{Overview of DSP Deployments Considered (Changes over the Baseline Marked in \textbf{bold})}
    \label{tab:overview:dsp}
    \setlength{\extrarowheight}{2pt}
    \begin{tabularx}{\textwidth}{lllllll}
        \toprule
                           & GCP (Baseline)         & GCP Pub/Sub            & AWS                         & GCP Samza              & GCP Dataflow            & GCP Autopilot          \\
        \midrule
        Cloud Provider     & GCP                    & GCP                    & \textbf{AWS}                & GCP                    & GCP                     & GCP                    \\
        Kubernetes Service & GKE                    & GKE                    & \textbf{EKS}                & GKE                    & ---                     & \textbf{GKE Autopilot} \\
        VM Instance Type   & \texttt{e2-standard-4} & \texttt{e2-standard-4} & \textbf{\texttt{m5.xlarge}} & \texttt{e2-standard-4} & \texttt{e2-standard-4}  & ---                    \\
        Streaming Engine   & Apache Flink           & Apache Flink           & Apache Flink                & \textbf{Apache Samza}  & \textbf{Cloud Dataflow} & Apache Flink           \\
        Transport          & HTTP \& Kafka          & \textbf{Cloud Pub/Sub} & HTTP \& Kafka               & HTTP \& Kafka          & \textbf{Cloud Pub/Sub}  & HTTP \& Kafka          \\
        Database           & Cloud Firestore        & Cloud Firestore        & \textbf{AWS DynamoDB}       & Cloud Firestore        & Cloud Firestore         & Cloud Firestore        \\
        \bottomrule
    \end{tabularx}
\end{table*}

\begin{table*}
    \centering
    \caption{Overview of FaaS Deployments Considered (Changes over the Baseline Marked in \textbf{bold})}
    \label{tab:overview:faas}
    \begin{tabularx}{0.8\textwidth}{llllll}
        \toprule
                        & GCP (Baseline)  & GCP Pub/Sub            & AWS                   & GCP Go          & GCP NodeJS      \\
        \midrule
        Cloud Provider  & GCP             & GCP                    & \textbf{AWS}          & GCP             & GCP             \\
        FaaS Engine     & Cloud Functions & Cloud Functions        & \textbf{AWS} Lambda   & Cloud Functions & Cloud Functions \\
        Function Memory & 256\,MB         & 256\,MB                & 256\,MB               & 256\,MB         & 256\,MB         \\
        Language        & Java            & Java                   & Java                  & \textbf{Go}     & \textbf{NodeJS} \\
        Transport       & HTTP            & \textbf{Cloud Pub/Sub} & HTTP                  & HTTP            & HTTP            \\
        Database        & Cloud Firestore & Cloud Firestore        & \textbf{AWS DynamoDB} & Cloud Firestore & Cloud Firestore \\
        \bottomrule
    \end{tabularx}
\end{table*}

As a second use-case, we chose a sliding window data aggregation (\emph{UC2}).
Such aggregations are used in many scenarios, e.g., to derive a smoothed trend.
As illustrated in \cref{fig:uc3}, incoming data is first windowed in a sliding window.
Data within a window is then aggregated by computing summary statistics, yielding a moving aggregation.
Results of this windowed aggregation can then be used in further data processing.
In this use-case, the windowing of data requires application state.

\subsection{Benchmark Methodology}

Executing the benchmark for a platform entails threes steps:

\begin{enumerate}
    \item An application containing the two use-cases is implemented for the chosen platform.
    \item For different rates of data ingress, i.e., workload levels, load is generated against the application.
    \item The total cost of running the application is measured for the duration of processing a constant data rate.
\end{enumerate}

\subsubsection*{\textbf{Application Implementation}}
Our benchmark first requires that the two use-cases are implemented for a chosen SUT, such as a DSP in a specific cloud environment or a FaaS offering.
Although we present some implementations in \cref{sec:experiments}, we cannot provide a generic, ready-to-use implementation for \textit{all} possible SUTs, as implementation details are highly SUT-specific.
The application implementation should also conform to any best practices for the chosen SUT to support a fair comparison~\cite{book_cloud_service_benchmarking}.

The benchmark is not restricted to evaluating only the difference between DSP and FaaS, but can also be used to evaluate other scenarios:
Users might, e.g., use the benchmark to compare two different stream processing engines, compare the same engine deployed on different cloud providers, or compare the same FaaS functions using different event triggers.
We explore such options in our experiments in \cref{sec:experiments}.

\subsubsection*{\textbf{Load Generation}}
Load is generated through a dedicated load generator deployed within the same cloud datacenter as the SUT using the Theodolite load generators.
The load generators emulate a number of sensors that send data in a fixed interval in an open workload model~\cite{book_cloud_service_benchmarking}, i.e., requests are non-blocking.
By varying the number of sensors that are emulated by the load generator, we can achieve cost estimates for varying request loads.
To simplify our cost calculations, we set the fixed interval at one second so that, e.g., 500 emulated sensors lead to a load of 500 requests/s.

A constant arrival rate does not necessarily reflect real world data ingress patterns.
However, the goal of our benchmark is not to measure scalability or elasticity of a given platform but rather to explore the cost of operating an application for a given data rate that may reflect an average rate over time.

\subsubsection*{\textbf{Cost Measurement}}
The result of our benchmark is an hourly cost estimate for the implementation of a use-case for a given level of constant load.
To yield such an estimate, we can leverage different kinds of information provided by cloud platforms or measurements.
For experiments on FaaS platforms with pay-per-request pricing models, cost estimates can be derived by extrapolating from small-scale environments, as the cost can be expected to scale linearly with the number of requests for current cloud pricing models.
Additionally, any costs for database reads and writes can be derived by tracking database access and calculating the resulting cost based on per-request cost of the database system used.
To achieve a cost estimate for a DSP deployment, we benchmark multiple infrastructure configurations until the least expensive deployment, which can still handle the configured load without violating specified service level objectives (SLOs), is found~\cite{EMSE2022, Brataas2017}.
Specifically, our SLO demands that the event consumer lag does not increase substantially~\cite{Henning2021}, meaning that events are processed at the same or higher rate of event ingress.
We evaluate this using monitoring data provided by the SUT and the cloud providers that are automatically analyzed by the Theodolite framework~\cite{TheodoliteDemoIC2E}.

%% file: sections/4_experiments.tex
\section{Experiments}
\label{sec:experiments}

\begin{figure*}
    \centering
    \subfloat[UC1 as FaaS Implementation\label{fig:deployment:faas-uc1-http}]{
        \includegraphics[width=0.99\columnwidth]{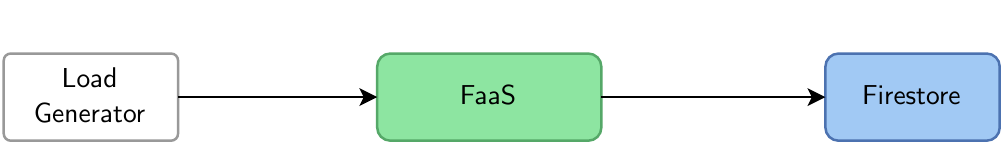}
    }
    \hfill
    \subfloat[UC1 as Streaming Implementation\label{fig:deployment:streaming-uc1-http}]{
        \includegraphics[width=0.99\columnwidth]{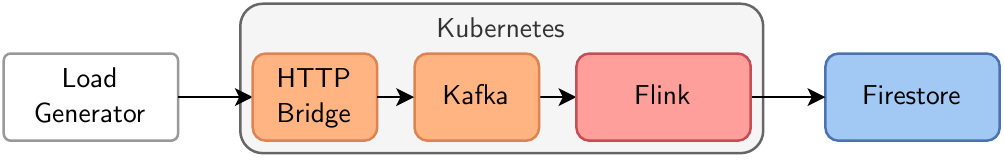}
    }
    \vfill
    \subfloat[UC2 as FaaS Implementation\label{fig:deployment:faas-uc2-http}]{
        \includegraphics[width=0.99\columnwidth]{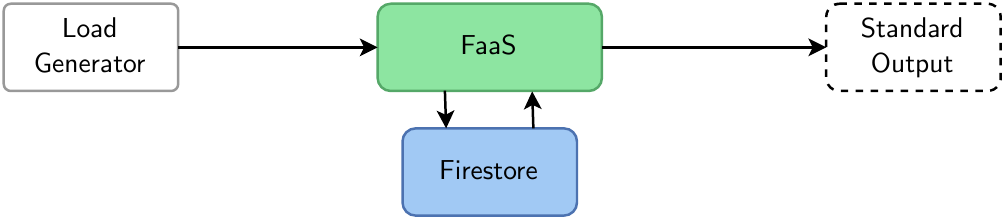}
    }
    \hfill
    \subfloat[UC2 as Streaming Implementation\label{fig:deployment:streaming-uc2-http}]{
        \includegraphics[width=0.99\columnwidth]{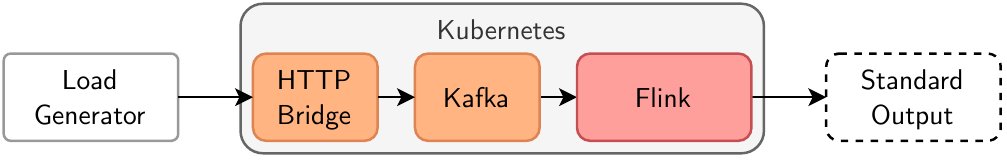}
    }

    \caption{Implementations in our baseline benchmarks: To aggregate data across multiple events, the FaaS implementation is connected to a Firestore database to persist state. As Apache Beam running on top of Apache Flink cannot process HTTP requests directly, we add an HTTP bridge and Apache Kafka.}
    \label{fig:deployment}
\end{figure*}

In this section, we present an extensive evaluation of different cloud event processing deployments.
After an initial comparison of DSP and FaaS (\cref{experiment:baseline}), we use our benchmark to explore the parameter space.
Specifically, we evaluate the impact of chosen event passing paradigm (\cref{experiment:pubsub}), cloud service provider (\cref{experiment:faas_vendors,experiment:kubernetes_vendors}), FaaS runtime environment (\cref{experiment:faas_runtimes}), DSP engine choice (\cref{experiment:streaming_engines}), a serverless DSP offering (\cref{experiment:serverless_streaming}), and a managed Kubernetes service (\cref{experiment:serverless_kubernetes}).
An overview of our experiment setups for DSP and FaaS is given in \cref{tab:overview:dsp,tab:overview:faas}, respectively.

\subsection{Baseline: Cloud Stream Processing and Functions}
\label{experiment:baseline}

As our baseline, we compare Google Cloud Functions and Apache Flink, running Apache Beam pipelines on Google Kubernetes Engine (GKE).

\subsubsection*{\textbf{Implementation}}
In the stateless storage use-case (UC1), client events are sent over HTTP and stored in Google Cloud Firestore (see \cref{fig:deployment:faas-uc1-http}).
We choose Firestore for its pay-as-you-go model that fits the serverless pricing model.
As necessary for Apache Flink, HTTP events are enqueued in Apache Kafka by a middleware prior to processing (see \cref{fig:deployment:streaming-uc1-http}).
Cloud Functions, on the other hand, can directly expose an HTTP endpoint.

The stateful windowed aggregation application (UC2) also receives events over HTTP, but results are emitted to the output log of the respective platform.
In a real application, a further stateless operation such as UC1 might be performed afterwards, yet our goal here is to study the stateful operator in isolation.
For our implementation with Flink, we use the built-in window aggregation mechanisms with RocksDB as state backend (see \cref{fig:deployment:streaming-uc2-http}).
To support stateful windowed aggregation on stateless functions, we store intermediate window state in a Google Cloud Firestore collection for each window (see \cref{fig:deployment:faas-uc2-http}).
Both implementations are configured to aggregate data over windows of 30~seconds, with a new window starting every 3~seconds. This results in 10~windows per emulated sensor that are maintained in parallel.

As Apache Flink and its operators are implemented in Java, we also use the Java~11 runtime for our cloud functions to account for effects caused by programming language or runtime.
We set the function memory to 256\,MB, which is the smallest amount that can support a function execution without running into memory errors.
This also limits our per-function compute resources to 0.1667\,vCPU.

For our streaming implementation, we deploy Flink in a GKE cluster with different numbers of \texttt{e2-standard-4} virtual machines.
The overall deployment consists of one coordinating Flink \textit{jobmanager}, varying numbers of Flink \textit{taskmanagers}, a three-node Apache Kafka cluster, a component redirecting incoming HTTP requests to Kafka as well as some additional components for monitoring and cluster management. To ensure a reasonable degree of fault tolerance, Flink is configured with a 30-second checkpointing interval and each Kafka partition is replicated across three brokers.

All experiments are conducted in the \texttt{europe-west-3} (Frankfurt) Google Cloud region, with the load generators deployed on \texttt{e2-highcpu-4} virtual machines on Google Compute Engine in the same region.

\subsubsection*{\textbf{Results}}

\begin{figure}
    \centering
    \subfloat[UC1 Costs\label{fig:baseline:uc1}]{
        \includegraphics[width=0.99\columnwidth]{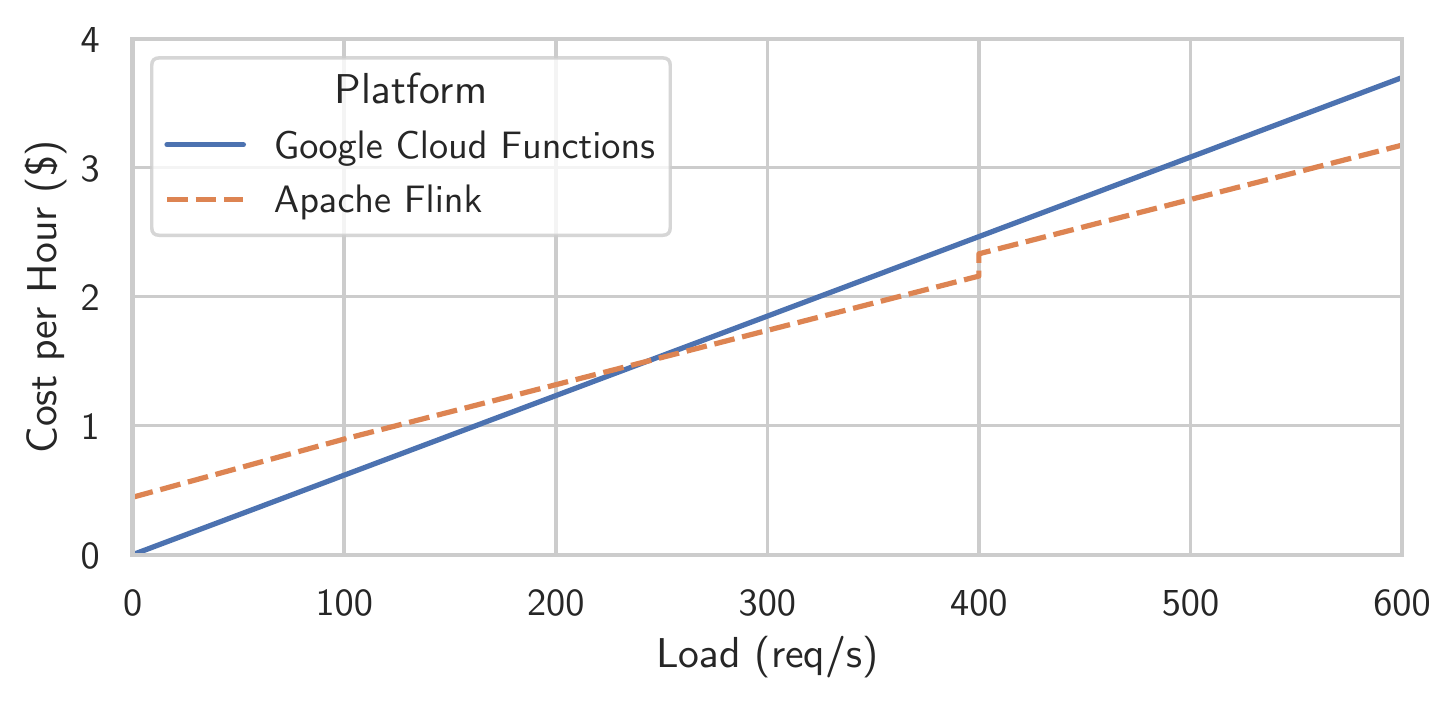}
    }
    \vfill
    \subfloat[UC2 Costs\label{fig:baseline:uc2}]{
        \includegraphics[width=0.99\columnwidth]{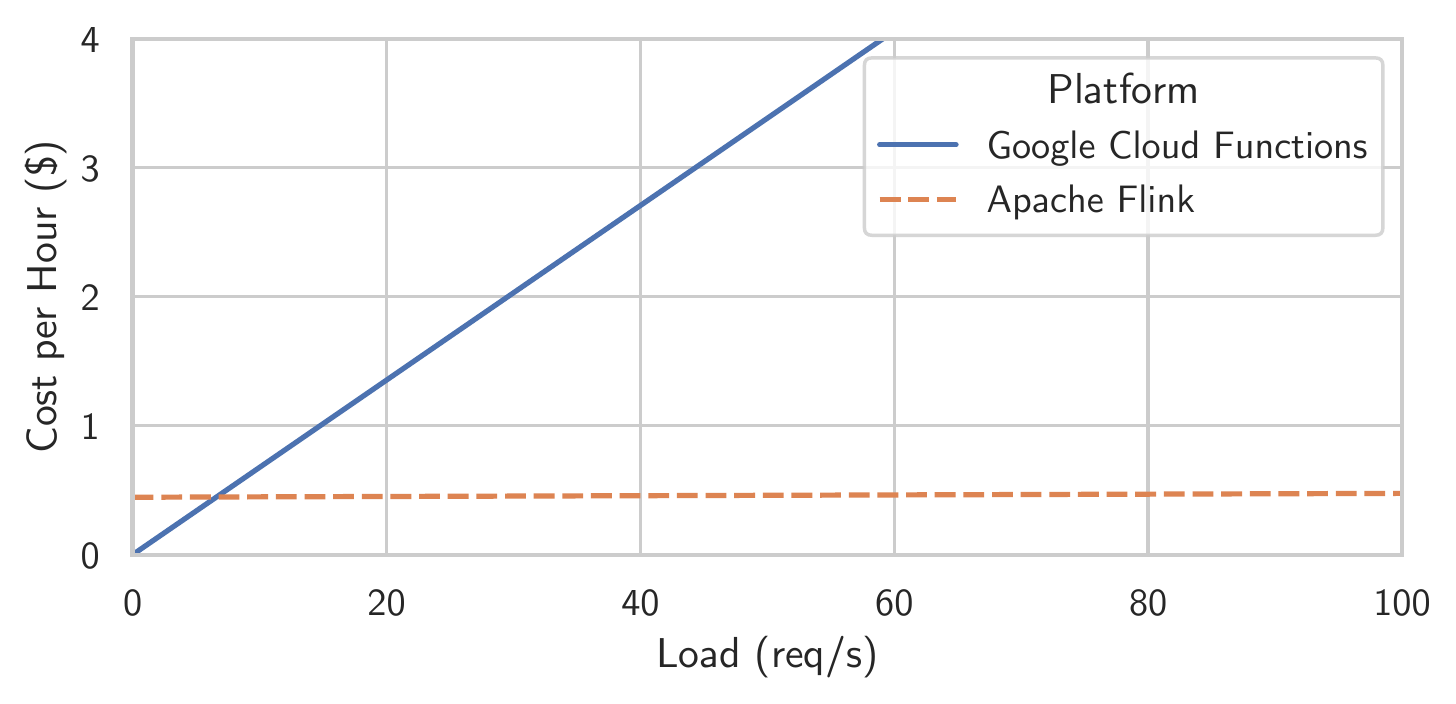}
    }
    \caption{The cost benchmark results of our baseline comparison of Apache Flink and Google Cloud Functions show how application costs scale with request load. The overhead of operating a Kubernetes cluster for Apache Flink leads to higher costs compared to Cloud Functions at lower request loads. The request rate at which Cloud Functions become less economical than stream processing with Flink depends on the type of function: 200~req/s for UC1 and 5~req/s for UC2.}
    \label{fig:baseline}
\end{figure}

\begin{figure}
    \centering
    \includegraphics[width=0.99\columnwidth]{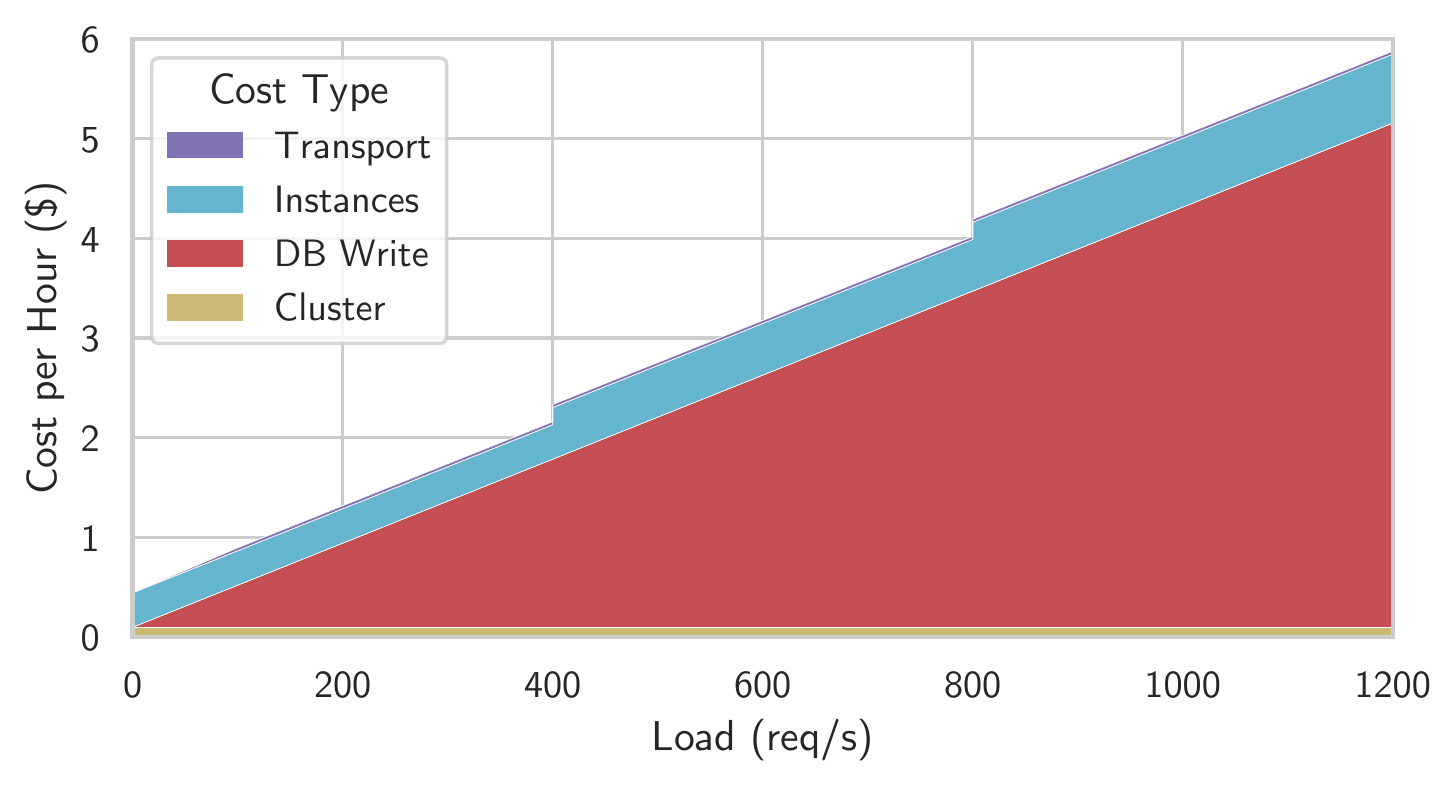}
    \caption{The cost breakdown of our baseline evaluation of UC1 with Apache Flink shows that total costs are composed of fixed costs (Kubernetes cluster and HTTP load balancer), costs per request (database writes), and costs increasing in batches (Kubernetes cluster nodes).}
    \label{fig:streaming-share-time:uc1}
\end{figure}

We show the results of our baseline evaluation in \cref{fig:baseline}.
For the application that we consider, costs scale linearly with request loads, yet at different rates.
This is expected for functions, which are billed by request and where requests can be processed independently.
In essence, FaaS is variable cost only.
In stream processing, we instead observe a pattern of \emph{steps}, which can be seen in \cref{fig:baseline:uc1} (and more pronounced at a larger scale in \cref{fig:streaming-platforms:uc2}).
This is a result of a more coarsely grained allocation of resources, i.e., servers that need to be added to the cluster.
Additionally, there is a minimum cost of running the cluster, which is the cost of a single server, a fixed rate for managing the Kubernetes cluster, and cost for the necessary load balancer.
Overall, this means that DSP costs here are a combination of fixed cost, variable cost per request, and variable cost which need to be added in batches, as shown in \cref{fig:streaming-share-time:uc1}.
This leads to the intersection of function and cluster costs at a specific request level (200~req/s for UC1 and 5~req/s for UC2):
At a request rate below this level, the fixed cost of running a single-server cluster is higher than paying per request for FaaS functions.
Beyond this request rate, the overhead of operating full servers in a cluster is negligible compared to the premium of serverless functions.

Interestingly, the break-even point is at a higher load rate for the stateless UC1 than for the stateful aggregation in UC2.
For the cloud function implementation of UC2, the largest share of costs per request are caused by writes (62.2\%) and reads (20.8\%) to Cloud Firestore, as shown in \cref{fig:function-share}.
This database access is required to store intermediate state -- in our implementation, each window is stored as a database entry, leading to ten read and write requests for each function invocation.
In the streaming implementation, on the other hand, there is no such database access required since all state is maintained inside the Flink taskmanagers.

\subsubsection*{\textbf{Takeaway for Platform Choice}}
Our baseline experiments show that FaaS is an economical choice over DSP for stateless applications with low to medium event arrival rates, in our case from 0 to 200 req/s.
For stateful applications, where functions need to store intermediate state in a database, the cost of database access makes FaaS infeasible for anything but low-rate event processing.

\subsection{Impact of Pub/Sub in FaaS and Streaming}
\label{experiment:pubsub}

While we use HTTP endpoints for sensors in our baseline evaluation, this does not necessarily reflect all IoT environments, where data distribution paradigms such as publish/subscribe are more common~\cite{paper_hasenburg_broadcast_groups}.
We thus further quantify the impact of endpoint choice on DSP and FaaS costs.

\begin{figure*}
    \centering
    \subfloat[UC1\label{fig:function-share:uc1}]{
        \includegraphics[width=0.99\columnwidth]{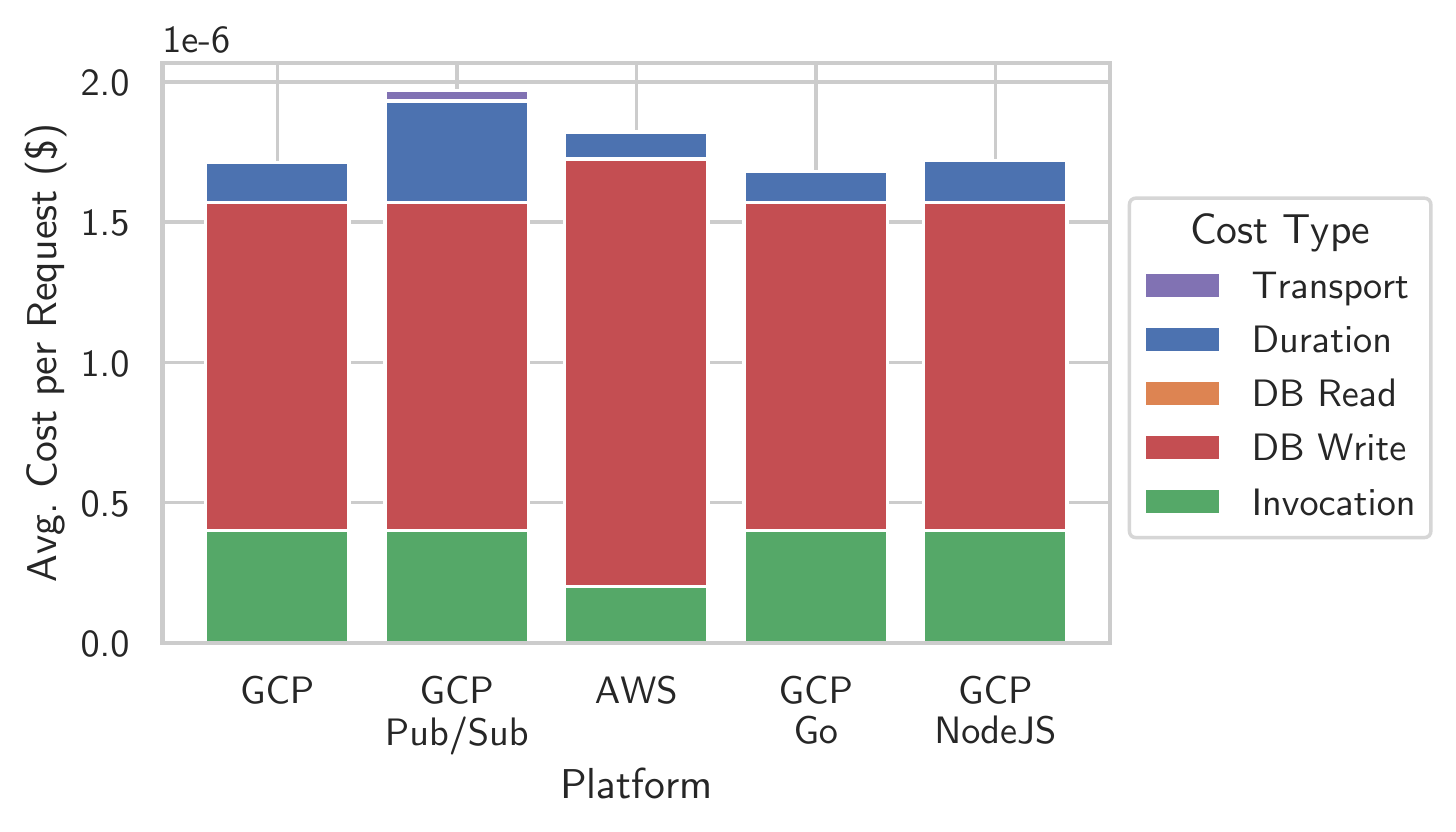}
    }
    \hfill
    \subfloat[UC2\label{fig:function-share:uc2}]{
        \includegraphics[width=0.99\columnwidth]{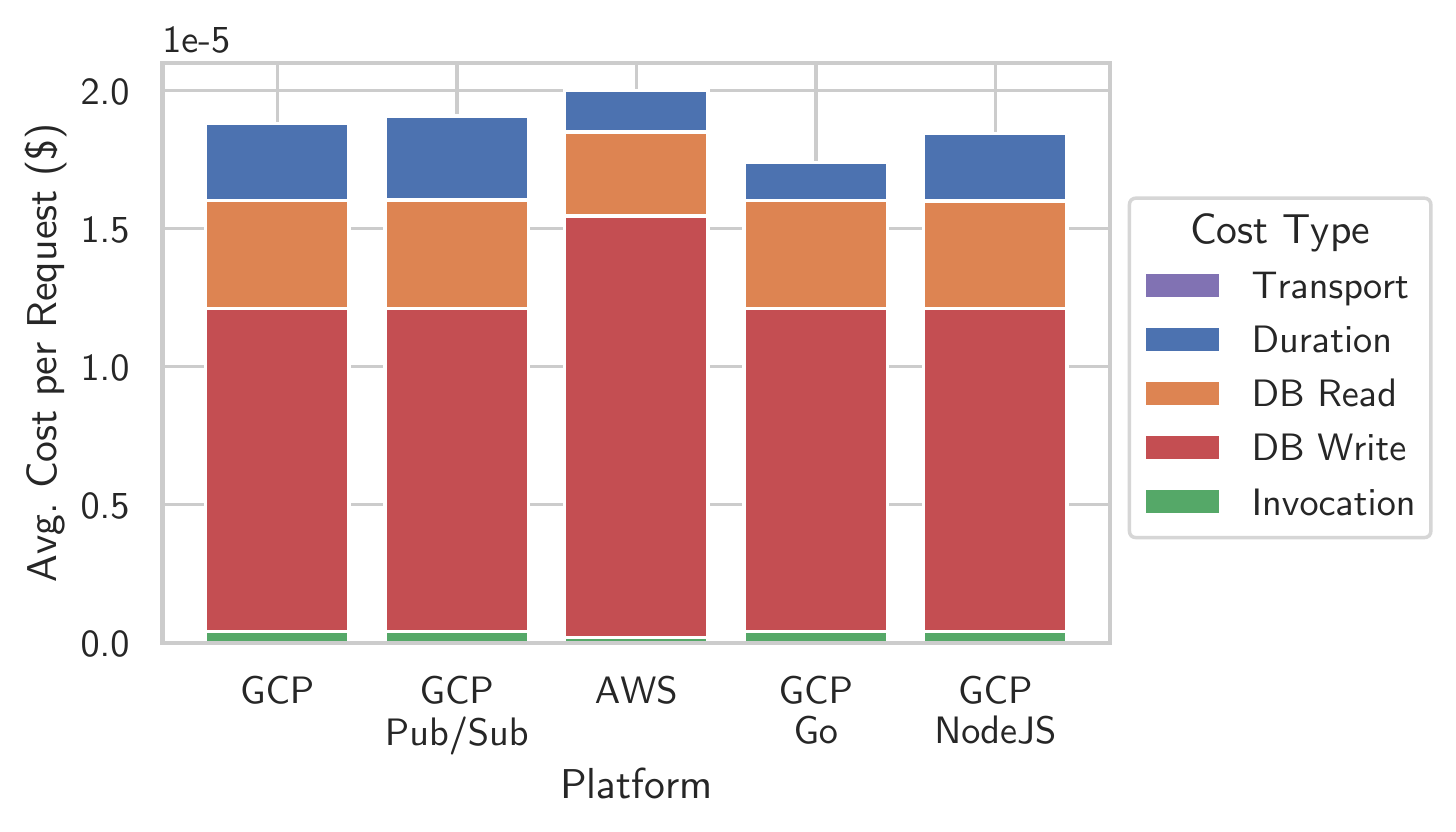}
    }
    \caption{\textbf{FaaS Cost per Request by Type:} Breaking down the costs per requests of our cloud function implementations of the two applications in our benchmarks, we see that database access is the major cost factor. While this does not impact UC1, where both the FaaS and DSP implementation write to Cloud Firestore and thus incur identical database access costs, storing intermediate state in UC2 accounts for 83.0\% of the total cost of operating the FaaS implementation. Neither the choice of Cloud Platform, of programming language, nor of endpoint change this result significantly: AWS Lambda is 6.4\% more expensive than our baseline as a result of increased DynamoDB access cost, while the choice of language runtime only changes function duration costs, which are marginal compared to Firestore access costs.}
    \label{fig:function-share}
\end{figure*}

\subsubsection*{\textbf{Implementation}}
We extend our baseline implementation with support for Google Cloud Pub/Sub\footnote{\url{https://cloud.google.com/pubsub/}}.
For our function implementation, this requires adding an event trigger and application logic for event parsing.
In our Apache Flink setup, we replace the previous HTTP middleware and the Apache Kafka deployment with a direct connection to Google Cloud Pub/Sub, using the \textit{PubSubIO} connectors provided by Apache Beam.
Instead of sending JSON objects as done with our HTTP implementation, we send binary encoded Apache Avro\footnote{\url{https://avro.apache.org/}} records via Pub/Sub.

\subsubsection*{\textbf{Results}}
As shown in \cref{fig:function-share}, using Cloud Pub/Sub has a noticeable effect on the execution duration of our FaaS implementations, especially in UC1, where processing costs increase by 154.6\%.
This effect is less pronounced for UC2, where duration increases by 8.6\%.
One possible explanation for this effect is an increased overhead caused by message parsing compared to HTTP, where request data is passed to our function directly as JSON rather than encoded.
However, due to the relatively high costs of database access, this has only a small impact on total costs (12.9\% increase for UC1 and 1.4\% increase for UC2).
At less than \$0.04 per 1,000,000 messages, the cost per Cloud Pub/Sub message is two orders of magnitude smaller than costs incurred by message processing.

\Cref{fig:streaming-platforms} shows how costs increase with increasing load when using Cloud Pub/Sub in our Apache Flink implementation. Pub/Sub introduces an additional cost factor to the overall deployment. These costs increase at a steeper rate than the costs for the Kubernetes cluster: While the share of Pub/Sub costs in total costs is 1.5\% for UC1 and 2.9\% for UC2 at a load intensity of 100 req/s, it grows to 2.6\% and 17.5\%, respectively, at a load of 1,000 req/s.
On the other hand, these additional costs are compensated by the slightly higher loads which Flink can process with Pub/Sub before requiring an additional virtual machine.
\Cref{fig:streaming-share} shows that, averaged over all evaluated load profiles, costs for processing messages from Pub/Sub are similar to redirecting HTTP requests via Kafka.

\subsubsection*{\textbf{Takeaway for Endpoint Choice}}

Our experiments show that there is no clear difference in costs when choosing Pub/Sub or HTTP, neither in DSP nor in FaaS. However, small savings are possible when using a transform method that simplifies processing. Hence, it does not seem to be reasonable to add a dedicated message transform layer just to save costs.

\subsection{Different FaaS Platforms}
\label{experiment:faas_vendors}

In our baseline FaaS evaluation, we use Google Cloud Functions, yet other cloud providers offer their own serverless platforms that may have different runtime behavior and pricing, impacting the cost results of our experiments.
In this experiment, we thus compare our Google Cloud Function implementation with an implementation on AWS Lambda.

\subsubsection*{\textbf{Implementation}}
We implement our benchmark for AWS Lambda with an AWS DynamoDB serverless database.
To ensure comparability, we use the Java~11 runtime and conduct our experiments in the \texttt{eu-central-1} (Frankfurt) region.
We again set the memory limit to 256\,MB.
Our load generator for this implementation runs in the same region on an \texttt{m5.xlarge} EC2 instance.

\subsubsection*{\textbf{Results}}

As we expect the costs for function execution to scale linearly with event arrival rate, we consider the average cost for individual function execution which we show in \cref{fig:function-share}.
The average cost per function execution is 6.4\% higher on AWS Lambda than on Google Cloud Functions for both applications, which is caused mainly by the more expensive database access in DynamoDB over Cloud Firestore.

\subsubsection*{\textbf{Takeaway for Cloud Provider Choice in FaaS}}
In our experiments, the choice of FaaS provider had only a limited impact on the total cost of execution, yet we see that the cost difference can depend on the type of application as applications using other cloud platform services may encounter significant costs (which may vary between providers).

\begin{figure}[b!]
    \centering
    \subfloat[UC1 Costs\label{fig:streaming-platforms:uc1}]{
        \includegraphics[width=0.99\columnwidth]{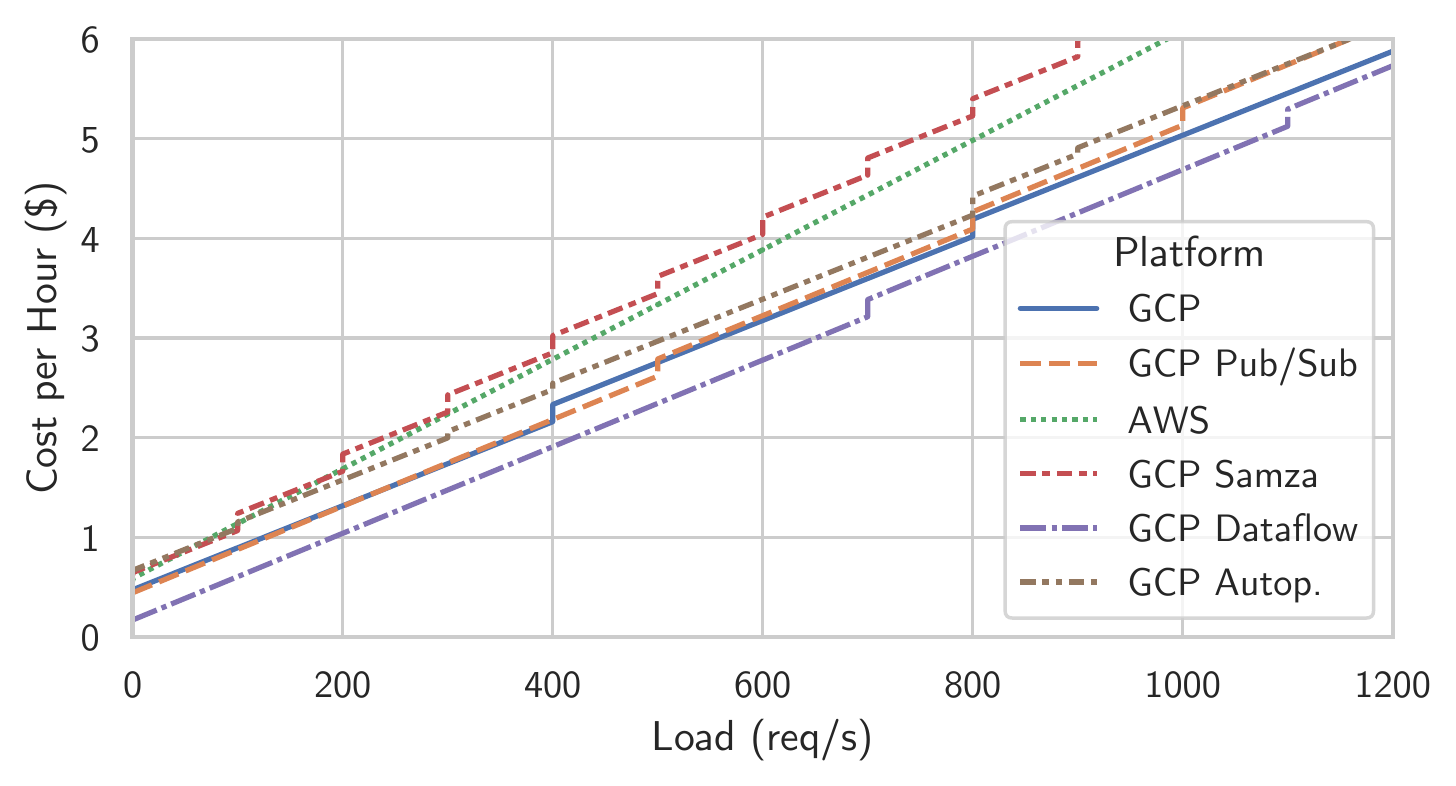}
    }
    \vfill
    \subfloat[UC2 Costs\label{fig:streaming-platforms:uc2}]{
        \includegraphics[width=0.99\columnwidth]{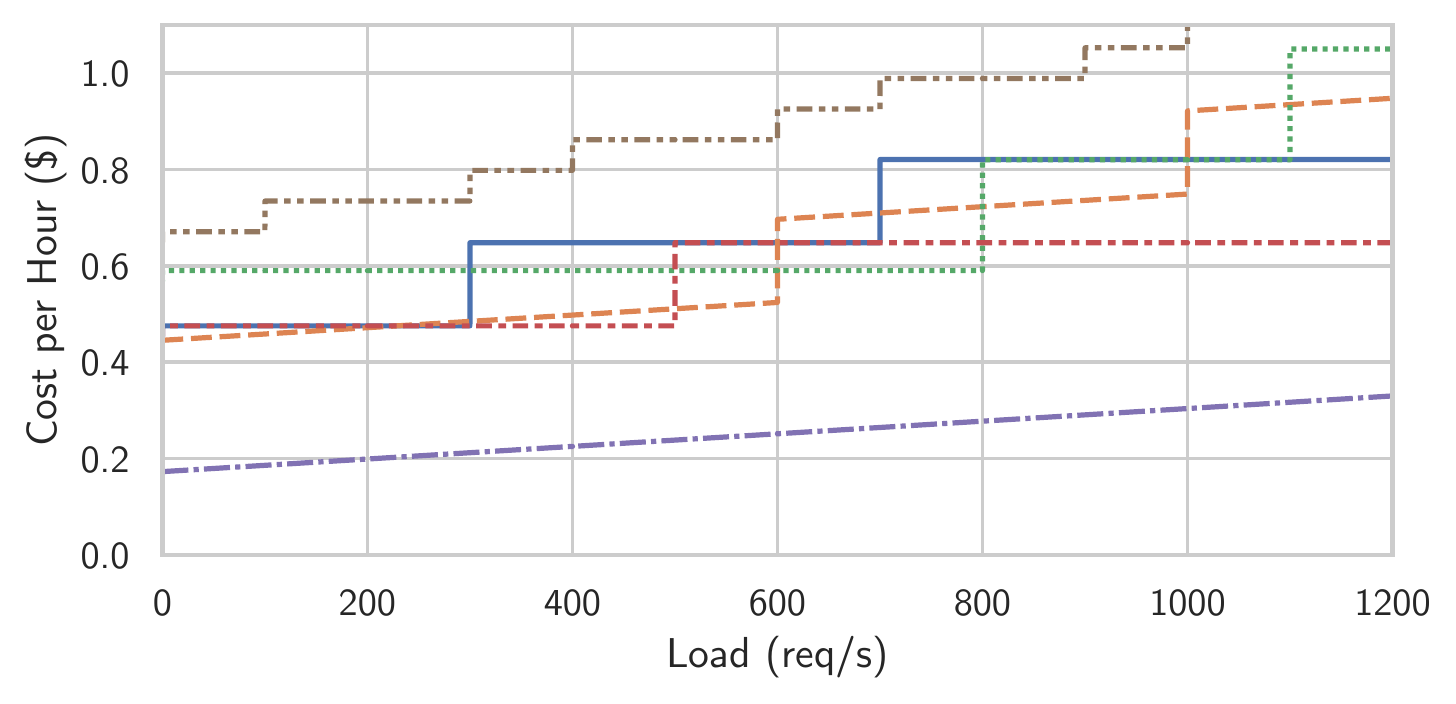}
    }
    \caption{Costs increase approximately linearly for all evaluated streaming deployments. However, Google Cloud Dataflow has considerably lower costs than the other streaming engines.}
    \label{fig:streaming-platforms}
\end{figure}

\begin{figure*}
    \centering
    \subfloat[UC1\label{fig:streaming-share:uc1}]{
        \includegraphics[width=0.99\columnwidth]{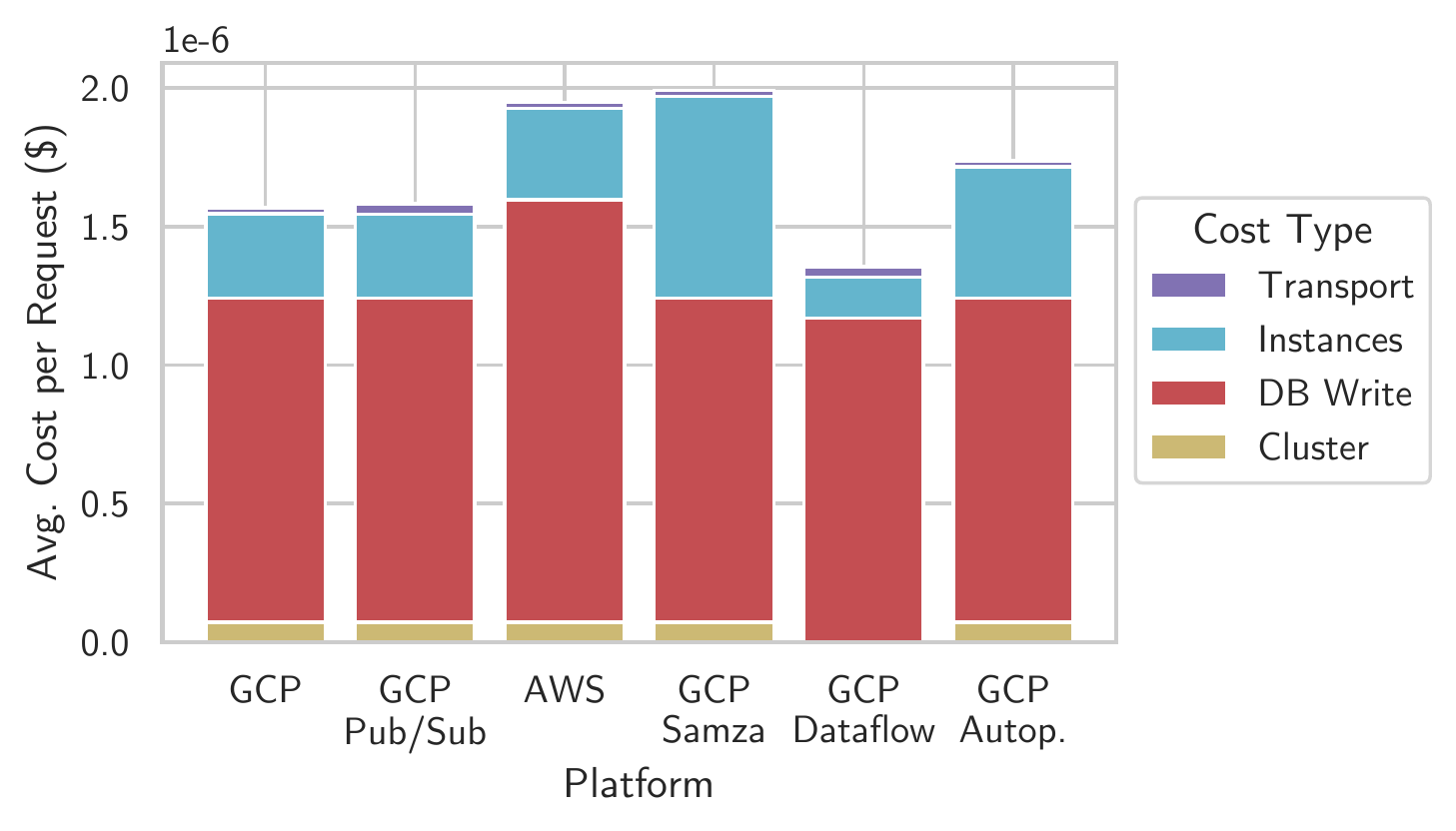}
    }
    \hfill
    \subfloat[UC2\label{fig:streaming-share:uc2}]{
        \includegraphics[width=0.99\columnwidth]{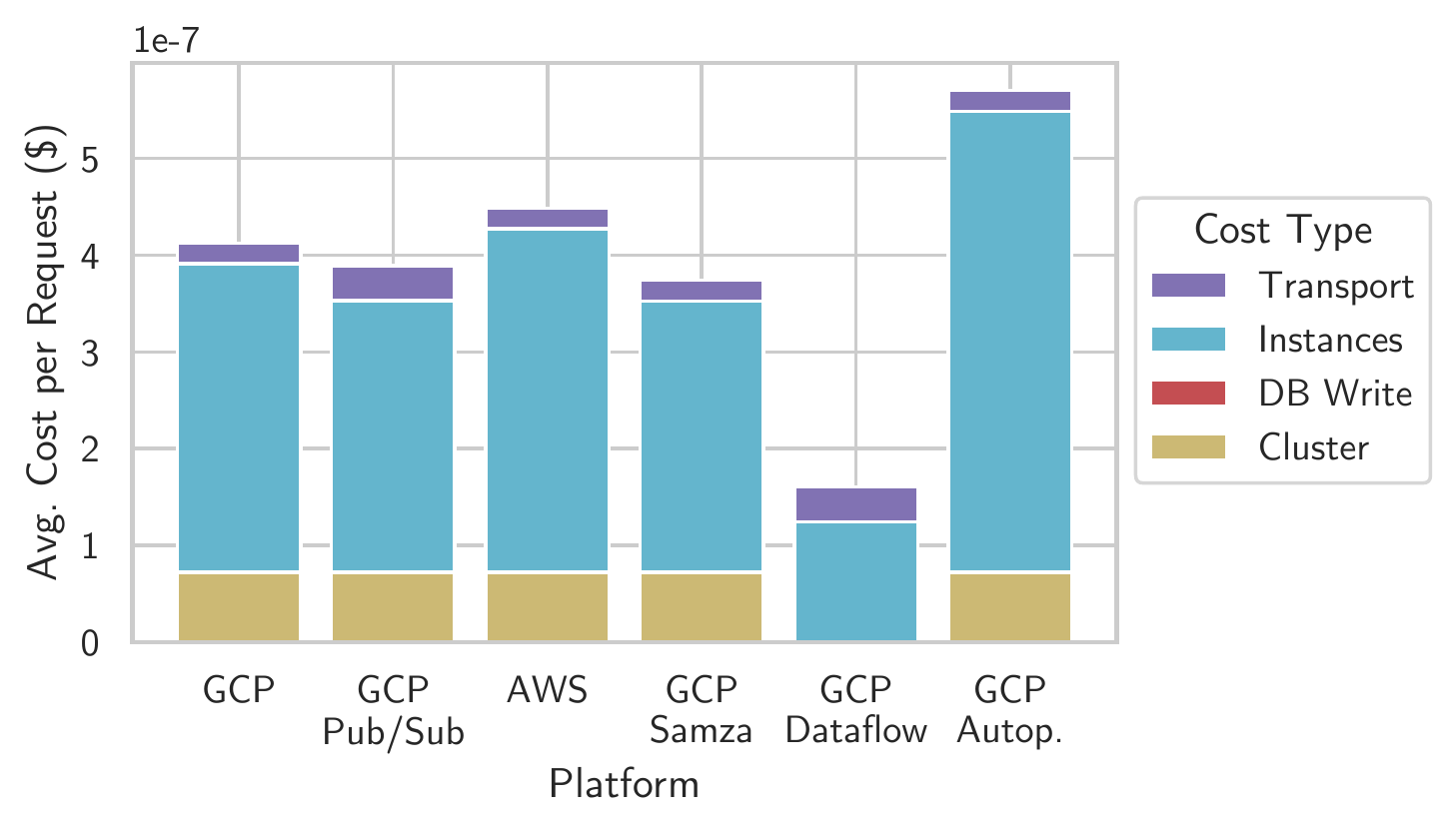}
    }
    \caption{\textbf{DSP Cost per Request by Type:} Averaging the cost per request over all evaluated load profiles, we see that, similar to FaaS, writing to a database is the largest cost factor for UC1 on all deployments. For UC2, costs are similar independent of the cloud provider, endpoint, and streaming engine, but instance costs are considerably lower for Dataflow and higher for GKE Autopilot.}
    \label{fig:streaming-share}
\end{figure*}

\subsection{Different Kubernetes Engines}
\label{experiment:kubernetes_vendors}

Similar to our evaluation of different FaaS Platforms, we also compare GKE and AWS Elastic Kubernetes Service (EKS).

\subsubsection*{\textbf{Implementation}}

Deployment descriptions for Kubernetes are largely platform independent, allowing us to almost use the same deployment with EKS as with GKE.
As in our evaluation of different FaaS Platforms, we write incoming events in our UC1 implementation to an AWS DynamoDB serverless database.
Both our EKS cluster and the load generator for this implementation use \texttt{m5.xlarge} EC2 instances, running in the \texttt{eu-central-1} (Frankfurt) region.

\subsubsection*{\textbf{Results}}

As shown in \cref{fig:streaming-platforms:uc1}, the costs for our UC1 deployment on EKS increase at a steeper rate than in the GKE deployment.
Averaged over all evaluated load profiles, EKS has 24.3\% higher costs than GKE as shown in \cref{fig:streaming-share:uc1}.
Interestingly, EKS has higher costs although the EKS deployment requires significantly less Flink taskmanager instances:
Loads up to 1,100 req/s can be processed by a single taskmanager, compared to 8~instances required in the GKE deployment.
However, higher costs per VM instance and especially higher costs per database write outweigh this superior performance.
As we do not see such a difference in resource usage for UC2, we conclude that either DynamoDB provides faster writes than Firestore or Beam's DynamoDB writer is more resource efficient than the Firestore writer.

In our implementation of the stateful application, we use only native Apache Beam functionality. As shown in \cref{fig:streaming-platforms:uc2}, costs increase in EKS at a similar rate as in GKE. Depending on the load intensity, at which VMs have to be added to the cluster, either GKE or EKS is cheaper. Averaged over all evaluated load profiles, EKS has 8.8\% higher costs than GKE (see \cref{fig:streaming-share:uc2}).
This is in accordance with the slightly higher costs per VM instance in AWS.

\subsubsection*{\textbf{Takeaway for Cloud Platform Choice in Stream Processing}}

Similar to our findings from evaluating different FaaS platforms, the choice of cloud infrastructure for running a DSP engine has a small but noteworthy impact on the total costs.
The discrepancy results mainly from different costs for cloud resources, which even outweigh significant performance gaps.

\subsection{Different Programming Languages in FaaS}
\label{experiment:faas_runtimes}

In our baseline FaaS evaluation, we use the Java~11 runtime in order to account for effects of programming language or runtime performance when comparing to Apache Flink.
Most modern FaaS platforms support a wider variety of runtimes, and the choice of language may have an indirect impact on execution cost when an implementation requires more resources or function executions take more time.

\subsubsection*{\textbf{Implementation}}
To quantify the effect of runtime choice, we implement our benchmark in Node.js and Go.
Node.js is one of the most popular choices for cloud functions, while Go is the only programming language supported by Google Cloud Functions that is compiled directly to machine code and may thus have the smallest performance overhead~\cite{9251194}.

\subsubsection*{\textbf{Results}}
As shown in \cref{fig:function-share}, the choice of programming language has only a small effect on the cost of function execution, with overall costs changing by -1.9\% and -7.5\% (Go) and 0.4\% and -1.9\% (Node.js) for UC1 and UC2, respectively.
Although the duration of a function execution changes by -22.7\% and -50.8\% for UC1 and UC2 with Go, the effect on costs is insignificant compared to costs for database access.
Surprisingly, the Node.js implementation is as efficient as our Java implementation.
This might be caused by a more mature and optimized execution environment in Google Cloud Functions, as Node.js is one of the most popular languages for FaaS functions.

\subsubsection*{\textbf{Takeaway for Language Choice in FaaS}}
As the majority of costs for the execution of a function are incurred by database access and not function duration, the choice of programming language has no considerable effect on the cost of our application.
For stateless applications without database access, and especially for more complex functions where the largest share of costs is incurred by execution duration rather than function invocation, comparing implementation runtimes may nevertheless be beneficial.

\subsection{Different Streaming Engines}
\label{experiment:streaming_engines}

We use Apache Flink for our baseline evaluation, which is a DSP engine originating in academia and extensively studied in research.
In this experiment, we compare this to Apache Samza, an open source DSP engine developed in industry at LinkedIn~\cite{Noghabi2017}.
Samza is built around similar concepts as Flink and can also be used to run Apache Beam pipelines.

\subsubsection*{\textbf{Implementation}}

Thanks to Apache Beam, we can use exactly the same implementation for Samza as we use for Flink.
In contrast to Flink, Samza does not need a dedicated coordinator, but instead uses our existing Kafka/ZooKeeper deployments for coordination among instances.

\subsubsection*{\textbf{Results}}

In case of the stateless application, we found that Samza has a significantly higher resource demand than Flink, causing higher costs as shown in \cref{fig:streaming-platforms:uc1}. As processing 300~req/s already requires 14~Samza instances, we extrapolated the costs for higher loads. We assume that this huge discrepancy is because we did not enable bundling, a Beam feature, which is used in Beam's \textit{FirestoreIO} to write multiple records as batch.
Bundling is disabled per default and its usage is not documented for Samza.

With the stateful application, Samza performs similar to Flink. As, however, Samza scales in smaller steps, the rather small load profiles studied here result in slightly lower costs for Samza as shown in \cref{fig:streaming-platforms:uc2}.

\subsubsection*{\textbf{Takeaway for Engine Choice}}

In general, different stream processing engines can be operated at similar costs. However, different feature sets and inappropriate configuration options might cause cost pitfalls, particularly when interacting with other cloud services.

\subsection{Serverless vs. Serverful Stream Processing}
\label{experiment:serverless_streaming}

In our baseline evaluation, we compare serverless FaaS implementations with streaming implementations running in Kubernetes. Major cloud vendors also provide managed streaming offerings, which run DSP pipelines on top of hosted stream processing engines.
While requiring the same development skills than with other DSP engines, serverless stream processing services can be considered an in-between of self-operated DSP engines and FaaS in terms of operational complexity.

\subsubsection*{\textbf{Implementation}}

To compare the costs of self-operating a DSP engine with a fully-managed one, we run our Apache Beam implementations on Google Cloud Dataflow with varying numbers \texttt{e2-standard-4} instances.
Similar to the other engines, Dataflow should be used with a durable data source instead of ingesting data directly via HTTP. As we consider using a serverless DSP service along with a self-operated Kafka cluster to be less realistic for real-world systems, we focus on processing data from Google Cloud Pub/Sub and use the Flink experiments with Pub/Sub as baseline.

\subsubsection*{\textbf{Results}}

As shown in \cref{fig:streaming-platforms}, Google Cloud Dataflow has significantly lower costs than our Apache Flink on Kubernetes deployment.
Averaged over all evaluated load profiles (see \cref{fig:streaming-share}), Dataflow has 85.6\% of the costs for operating Flink for UC1 and only 41.2\% for UC2.
This is primarily due to the massively reduced costs for the virtual machines as with Dataflow, fewer instances are required to process the same load, e.g., the stateful application can be run with a single VM at all tested load rates.
We observed that costs for Dataflow could be further reduced when using smaller instances such as \texttt{n1-standard-1} ones. %
Additionally, there are no general managing fees for Dataflow, while Google charges customers \$0.10 per hour for managing a Kubernetes cluster.
The impact of this fee on total costs decreases with increasing load (see \cref{fig:streaming-share-time:uc1}).
Since the largest cost driver in the stateless application are database writes, costs are reduced less than in the stateful application.
An in-depth analysis of resource efficiency advantages in Dataflow is beyond the scope of this work, but possible reasons are:

\begin{itemize}
    \item Dataflow might in general offer a better performance than other stream processing engines.
    \item Apache Beam might be optimized for Google Cloud Dataflow and, as shown in previous research \cite{Hesse2019}, Flink provides much better performance when running native Flink pipelines instead of using Beam.
    \item Flink's default configuration might not be optimal and additional tuning is required to reach comparable performance.
    \item Resource utilization when running Flink in small Kubernetes clusters might not be optimal. %
\end{itemize}

\subsubsection*{\textbf{Takeaway for Platform Choice}}

Processing event streams with Google Cloud Dataflow had significantly lower costs in our experiments compared to our Flink deployment.
Thus, serverless stream processing services can be a compelling alternative to running stream processing engines manually in Kubernetes, reducing both operational complexity and costs.

\subsection{Serverless vs. Serverful Kubernetes}
\label{experiment:serverless_kubernetes}

Recently, cloud providers started offering managed Kubernetes services, which charge users per container resource usage instead of for the underlying VM instances.
A prominent example for such a service is GKE Autopilot~\cite{cloudautopilot}. %

\subsubsection*{\textbf{Implementation}}

As autoscaling of the Kubernetes cluster takes a considerable amount of time, running dedicated experiments with GKE Autopilot is unpractical. However, we can get a reasonable cost approximation by using the results of our baseline evaluation, in which we determined the required number of Flink taskmanagers per load profile on a sufficiently dimensioned cluster. Total costs are then the costs for the taskmanagers, combined with the constant costs for other components such as Kafka, HTTP Bridge, or monitoring.

\subsubsection*{\textbf{Results}}

Independent of the load profile and the use case, GKE Autopilot has higher costs compared to GKE's default mode (see \cref{fig:streaming-platforms}). The relative cost difference appears to decrease with higher loads. This can be explained by a minimal cost per container that is charged independent of the actual resource usage.
Moreover, the cost difference is less pronounced in the stateless application, where costs are heavily influenced by database writes (see \cref{fig:streaming-share}).

\subsubsection*{\textbf{Takeaway for Engine Choice}}

While serverless Kubernetes offerings reduce the management burden, they also have higher cloud service costs. Nevertheless, costs for running self-operated DSP engines in a serverless Kubernetes cluster are still lower than for FaaS at medium and high loads.

%% file: sections/5_guidelines.tex
\section{Decision Guidelines}
\label{sec:guidelines}

\begin{figure}
    \centering
    \includegraphics[width=0.6\linewidth]{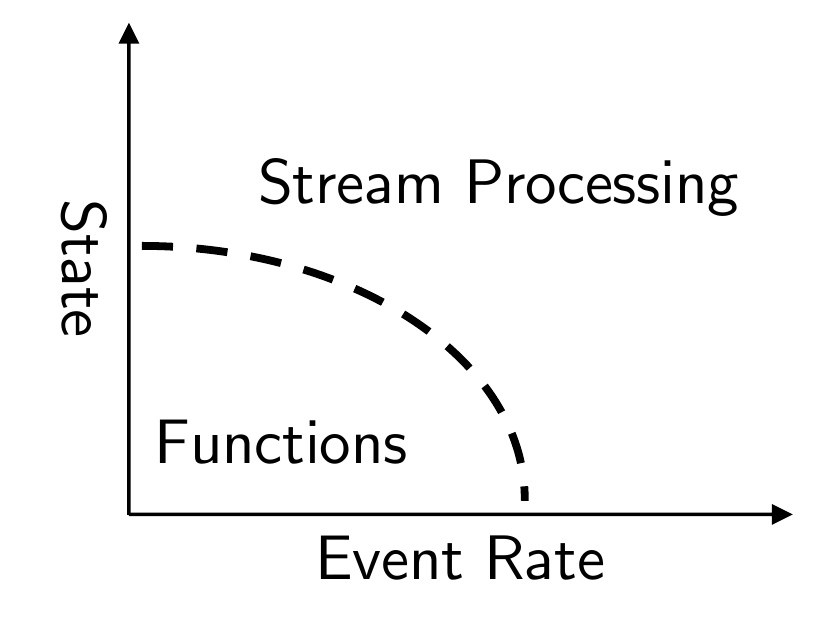}
    \caption{From a cost perspective, FaaS is the best choice for applications that require less state and process less events.}
    \label{fig:tradeoff}
\end{figure}

In our experiments, we have quantitatively evaluated the choice between functions and stream processing for cloud event processing and have explored the impact of choosing cloud providers, endpoints, programming languages, and platforms.
We see that the major influences on cost are the rate at which events arrive and the type of application.
As shown in \cref{fig:tradeoff}, FaaS is the economic choice for applications that manage little to no state and process events with low to medium arrival rates.
DSP is better suited for operations that require state, such as windowed aggregation, and for applications that process more events, i.e., on the order of thousands of events per second.

Beyond these considerations, we could not observe any considerable impact of other deployment parameters on costs.
The choice of a specific messaging paradigm, such as Pub/Sub or HTTP, should thus be based not on cost but on functional differences.
Similarly, the choice of cloud service provider did not influence costs significantly and might be influenced more by specific services that a provider offers.

%% file: sections/6_discussion.tex
\section{Limitations \& Future Research Directions}
\label{sec:discussion}

In our benchmarks and guidelines, we consider solely the cost incurred by cloud resources for different deployments of our applications.
Particularly, we did not try to quantify the ``human resource'' costs for implementing, operating, and maintaining a specific target design.
Beyond both cost types, there are other aspects that may influence the design of a cloud event processing application.
We discuss these perspectives here and derive avenues in which our work could be extended in the future.

\subsection{Non-Constant Workloads \& Elasticity}

In our benchmark experiments, we consider a constant event arrival rate as our goal is to measure deployment costs at a specific load.
In some domains, workloads may instead fluctuate, requiring elasticity from the processing application.
This elasticity is handled differently in DSP and FaaS:
As functions are stateless and can be scaled horizontally quickly, load peaks can be processed in real-time.
This will briefly increase costs for a FaaS deployment.
In DSP, such peaks may be handled by queuing events and processing them once load has reduced.
This does not require any additional infrastructure and hence does not incur additional costs as long as sufficient queue capacity exists.
Alternatively, infrastructure can be expanded easily by adding more compute nodes to the cluster.
Compared to FaaS platforms, such horizontal scaling is rather slow and will still require queuing.
Depending on the billing scheme of the runtime platform as well as the scale-in strategy, short load spikes can also mean that the DSP cluster is overprovisioned (and thus over-expensive) for some time after the load spike whereas FaaS providers pay the costs for keeping functions warm after a load spike.

\subsection{Stateful Functions}

Our experiments show that building a stateful processor with serverless functions leads to high costs incurred by database access used to persist state.
Recently, there have been some proposals to add mechanisms for stateful stream processing to function platforms, e.g.,~\cite{Akhter2019, Sreekanti2020, deHeus2021}.
These approaches typically include a dedicated datastore directly in the FaaS platform, which could reduce access costs.
However, public cloud vendors do not offer such services at this time, leaving engineers only the option of dedicated cloud datastores.
As an alternative, engineers might use an open source FaaS system and retrofit the ``sharding by key'' features to its load balancer and use local ephemeral storage for state.
This would require significant engineering and infrastructure management efforts, breaking the concept of ``serverless'' platforms.

\subsection{Lock-In Effects}

In addition to deployment costs, there are ``hidden'' costs to building cloud applications with managed services such as Kubernetes engines or FaaS platforms:
Lock-in effects increase the effort required to move between cloud vendors.
Such effects could also influence the decision between DSP and FaaS as paradigms for a cloud event processing application:
We were able to move our Apache Flink benchmark implementation from Google Kubernetes Engine to AWS EKS with little effort (\cref{experiment:kubernetes_vendors}) as both platforms understand similar Kubernetes application descriptions.
Porting our implementation from Google Cloud Functions to AWS Lambda (\cref{experiment:faas_vendors}), however, required changing the highly platform-specific function implementation almost completely.

\subsection{SLAs and SLOs}

A further factor that is beyond the scope of our benchmark is the influence of different service level agreements (SLA) and service level objectives (SLO) on the true cost of an application deployment.
For self-managed streaming applications in Kubernetes, only very basic SLAs are guaranteed by the cloud provider such as the availability of compute instances.
Application-level SLOs such as maximum latency must be monitored and managed by the operator.
As FaaS platforms are fully managed by the provider, they may provide further guarantees on availability.

\subsection{Tuning for Cost-efficiency}

Finding a cost-optimal configuration (e.g, machine type, cluster size, or stream processing engine settings) for a self-operated DSP deployment is a complex task, especially in comparison to FaaS.
This is even more important when comparing managed stream processing services against self-operated
ones and may also explain why we found Cloud Dataflow to be significantly less expensive than running Apache Flink.
We cannot exclude that Apache Flink can be tuned for better performance to achieve similar or better cost efficiency than FaaS for low event rates or than Google Cloud Dataflow in general. However, such performance tunings come at the cost of expert knowledge or extensive benchmarking.

%% file: sections/7_relwork.tex
\section{Related Work}
\label{sec:relwork}

Although including a cost model in cloud benchmarking studies is considered good scientific practice~\cite{Papadopoulos2021}, in existing benchmarking studies on FaaS~\cite{Kuhlenkamp2020,Copik2021,Ngo2022,paper_grambow_befaas,vanEyk2020} and DSP~\cite{Bordin2020,VanDongen2021,BDR2021}, cost evaluations can mainly be found for cloud functions, where the pay-per-execution pricing model has presented a significant paradigm shift.

\emph{LIBRA}~\cite{Raza2021-fj} is an approach to offload FaaS function invocations to self-managed function infrastructure to leverage economies of scale and decrease costs for FaaS applications.
Conversely, \emph{SplitServe}~\cite{Jain2020-mh} offloads latency-sensitive Apache Spark jobs to a FaaS platform to manage unexpected spikes in demand.
Chadha et al.~\cite{9582234} present a comprehensive evaluation of the impacts of runtime, region, and processor architecture choice on the performance and cost of compute-intensive functions on Google Cloud Functions.
Similarly, Eivy~\cite{7912239} gives an overview and discussion of cloud FaaS pricing and Cordingly et al.~\cite{9251165} introduce \emph{SAAF}, a cost and performance predictor for serverless functions.
In the context of DSP, Truong et al.~\cite{8814567} present a resource provisioning strategy that optimizes costs for cloud data processing and Bedini et al.~\cite{10.1145/2479871.2479895} show an approach to model the performance of the Apache Storm stream processing engine.
To the best of our knowledge, existing work has not compared FaaS and DSP to implement the same application.

Copik et al.~\cite{Copik2021} evaluate how Infrastructure-as-a-Service costs relate to FaaS costs, finding that IaaS provides better performance at lower costs if high utilization could be reached.
Similarly, M\"{u}ller et al.~\cite{Mueller2020} compare the costs of Query-as-a-Service systems with FaaS costs and show that \textit{cold data} can be requested significantly cheaper with FaaS.

Previous research comparing different stream processing systems focuses on self-operated, open source systems such as Apache Storm, Apache Flink, and Apache Spark and does not include cloud services for DSP~\cite{Karimov22018, VanDongen2020, Hesse2021, BDR2021}.
Akidau et al.~\cite{Akidau2021} present a performance comparison of Apache Flink and Google Cloud Dataflow on GCP.
These evaluations, however, do not focus on cloud infrastructure costs.

In previous work~\cite{paper_pfandzelter_functions_streams}, we have considered the choice between functions, stream processing, and batch processing for IoT data and event processing in the fog from a qualitative perspective and derived a set of best practices.
With a focus on cloud event processing in this paper, we have extended this with a quantitative evaluation focusing on the cost dimension.

%% file: sections/8_conclusion.tex
\section{Conclusions and Outlook}
\label{sec:conclusion}

In this paper, we took a cost perspective on cloud event processing.
We have presented a novel application-centric cost benchmark with event dataflows from an IIoT context that include both a stateless and a stateful job graph.
Further, we have used this benchmark to compare distributed stream processing and Functions-as-a-Service, today's most popular cloud event processing paradigms, and have explored the parameter space to evaluate which factors influence the cost of operating event processing applications in the cloud.
We found that in terms of pure costs for the cloud services, FaaS is superior for applications that are subject to small event rates and require small state. Once event rates increase and utilize at least a single DSP instance, DSP engines can be operated at lower costs.
This observation holds independently of the cloud provider and implementation technology.
However, when choosing among FaaS and streaming also ``hidden'' costs should be taken into account.
As part of this paper, we derived avenues for future work to quantify the costs for other workload scenarios, future cloud services, lock-in effects, SLO compliance, and cost tuning.